\documentclass[manuscript]{emulateapj}

\shorttitle{Block-induced complex structures building the flare-productive solar active region 12673}
\shortauthors{Yang et al.}

\journalinfo{Accepted for publication in ApJL}
\submitted{ApJL accepted 2017 October 17}

\begin{document}

\title{Block-induced complex structures building the flare-productive solar active region 12673}

\author{Shuhong Yang\altaffilmark{1,2}, Jun Zhang\altaffilmark{1,2},
Xiaoshuai Zhu\altaffilmark{3}, and Qiao Song\altaffilmark{4}}

\altaffiltext{1}{CAS Key Laboratory of Solar Activity, National
Astronomical Observatories, Chinese Academy of Sciences, Beijing
100012, China; shuhongyang@nao.cas.cn}

\altaffiltext{2}{College of Astronomy and Space Sciences, University
of Chinese Academy of Sciences, Beijing 100049, China}

\altaffiltext{3}{Max-Planck Institute for Solar System Research, G\"{o}ttingen 37077, Germany}

\altaffiltext{4}{Key Laboratory of Space Weather, National Center for Space Weather, China Meteorological Administration, Beijing 100081}

\begin{abstract}
  Solar active region (AR) 12673 produced 4 X-class, 27 M-class, and numerous lower class flares during its passage across the visible solar disk in September 2017. Our study is to answer the questions why this AR was so flare-productive and how the X9.3 flare, the largest one of the last decade, took place. We find that there was a sunspot in the initial several days, and then two bipolar regions emerged  nearby it successively. Due to the standing of the pre-existing sunspot, the movement of the bipoles was blocked, while the pre-existing sunspot maintained its quasi-circular shaped umbra only with the disappearance of a part of penumbra. Thus, the bipolar patches were significantly distorted, and the opposite polarities formed two semi-circular shaped structures. After that, two sequences of new bipolar regions emerged within the narrow semi-circular zone, and the bipolar patches separated along the curved channel. The new bipoles sheared and interacted with the previous ones, forming a complex topological system, during which numerous flares occurred. At the highly sheared region, a great deal of free energy was accumulated. On September 6, one negative patch near the polarity inversion line began to rapidly rotate and shear with the surrounding positive fields, and consequently the X9.3 flare erupted. Our results reveal that the block-induced complex structures built the flare-productive AR and the X9.3 flare was triggered by an erupting filament due to the kink instability. To better illustrate this process, a block-induced eruption model is proposed for the first time.

\end{abstract}


\section{INTRODUCTION}

Solar flares are powerful energy release phenomena on the Sun, and magnetic reconnection is deemed to be the most effective mechanism of the rapid energy release (e.g., Sturrock 1966; Masuda et al. 1994; Yang et al. 2014). Magnetic flux emergence and cancellation were found to be favorable for flare occurrence and coronal mass ejection (CME) onset (Wang \& Shi 1993; Zirin \& Wang 1993; Schmieder et al. 1997; Chen \& Shibata 2000; Zhang et al. 2001; Sterling et al. 2010; Louis et al. 2015). Eruptions can be triggered through the reconnection between the emerging flux and the pre-existing flux. In a statistical study of 77 X-class and M-class flares by Burtseva \& Petrie (2013), the results revealed the importance of flux cancellation in triggering the flares. At the polarity inversion lines (PILs), the magnetic gradients are steep and the horizontal fields are highly sheared. Since the stressed magnetic fields contain a great deal of free energy, large flares tend to take place in the vicinity of the PILs (Sun et al. 2012; Song et al. 2013). The rapid shear flows in the photosphere are thought to play an crucial role in the buildup of free energy to power solar flares (Harvey \& Harvey 1976; Meunier \& Kosovichev 2003; Yang et al. 2004). In a study of Shimizu et al. (2014), strong flows along the flaring PIL of an X5.4 flare were observed before and after the flare, which were argued to shear the magnetic fields and increase the free energy to power the major flare. Besides the photospheric shear flows, sunspot rotation is also an important mechanism for triggering major flares (Brown et al. 2003; Zhang et al. 2007; Yan et al. 2009; Vemareddy et al. 2016).

Solar active region (AR) 12673 erupted with a series of flares, including 4 X-class and 27 M-class, from Sept. 4 to Sept. 10, 2017. The X9.3 flare on Sept. 6 is the largest one in Solar Cycle 24. One may be interested in the questions: (1) Why was this AR so flare-productive? (2) How did the X9.3 flare occur? The present paper is dedicated to answering these two questions.

\section{OBSERVATIONS AND DATA ANALYSIS}

The Helioseismic and Magnetic Imager (HMI; Scherrer et al. 2012; Schou et al. 2012) on board the \emph{Solar Dynamics Observatory} (\emph{SDO}; Pesnell et al. 2012) provides the full-disk line-of-sight (LOS) magnetograms and intensity maps with the pixel size of 0{\arcsec}.5 and the cadence of 45 seconds. The Atmospheric Imaging Assembly (AIA; Lemen et al. 2012) on board the \emph{SDO} observes the Sun in ten (E)UV channels with the spatial sampling of 0{\arcsec}.6 pixel$^{-1}$ and the cadence of (12)24 seconds. We use the HMI LOS magnetograms and intensity maps and AIA multi-wavelength images from Sept. 1 to Sept. 10 with a 15 min cadence, the data on Sept. 6 with a 3 min cadence, and the two hour data around the X9.3 flare with full cadence. All of them are calibrated to Level 2 using the standard routine under the Solar Software package. Then the data from Sept. 1 to Sept. 7 are differentially rotated to the reference time of 20:00 UT on Sept. 3. In addition, we also employ the \emph{Geostationary Operational Environmental Satellite} (\emph{GOES}) data with the 1 min cadence to examine the variation of soft X-ray 1-8 {\AA} flux.

\emph{SDO}/HMI also measures the full-disk vector magnetic fields with the spatial sampling of 0.$\arcsec$5 pixel$^{-1}$, and the inverted vector magnetograms using VFISV inversion code (Borrero et al. 2011) have a cadence of 12 min. We adopt two full-disk disambiguated magnetograms observed at 12:00 UT on Sept. 3 and at 09:48 UT on Sept. 6. Then the vector magnetic field at each pixel in the image plane is transformed to the heliographic components with the formulae given by Gary \& Hagyard (1990). Moreover, we perform the geometric mapping of the magnetograms into the heliographic coordinates. Based on the photospheric vector magnetograms, we reconstruct the coronal structures using the nonlinear force-free field (NLFFF) modeling (Wheatland et al. 2000; Wiegelmann 2004). Before the extrapolation, the magnetograms are preprocessed to best suit the force-free conditions (Wiegelmann et al. 2006). The extrapolation is performed in the cubic box of 512$\times$512$\times$256 uniformly spaced grid points with $\Delta x= \Delta y=\Delta z=0.$\arcsec$5$. AR 12673 is located at the center of the photospheric boundary. Furthermore, we use the code developed by Liu et al. (2016) to calculate the squashing factor $Q$ (D{\'e}moulin et al. 1996; Titov et al. 2002) and twist number $\mathcal{T}_w$ (Berger \& Prior 2006) of the reconstructed magnetic field.

\section{RESULTS}

Beginning on Sept. 4, AR 12673 erupted a series of flares during its passage across the visible solar disk. Table 1 shows the information of all the 4 X-class and 27 M-class flares\footnote[1]{Based on the data from NOAA (ftp://ftp.swpc.noaa.gov/pub/warehouse/2017/2017\_events/)} in AR 12673. For each flare, the mean duration is determined from the start to end time of the flare. The average duration of the X-class flares is about 32 min, longer than that (26 min) of the M-class flares. Among these flares, the X9.3 flare occurred on Sept. 6 is the largest one  since 2005.

In the initial two days of Sept., there was only one sunspot with positive polarity within AR 12673 (see Figure 1(a)). Then one bipolar region (labeled with ``Bipole A") emerged at the south-east of the pre-existing spot (panel (b)), followed by the successive emergence of another bipolar region (``Bipole B") at the north-east (panel (c)). The negative polarity patches of the bipoles moved east-ward and the positive ones moved west-ward (see Movie1). Due to the standing of the pre-existing spot, the west-ward motion of the positive patches was blocked. During this process, the pre-existing spot maintained its quasi-circle shape with its location almost no dramatic change. However, the left part of the penumbra disappeared due to the conflict with the nearby moving patches, as denoted by the arrows in panels (a) and (d). Thus, the bipoles were significantly distorted and elongated around the pre-existing spot, and their positive and negative polarities formed two semi-circular structures (panel (d)). After that, two sequences of new bipolar regions (labeled with ``Bipole C" and ``Bipole D") emerged within the narrow zone between the semi-circular structures (panel (e)). The newly emerging bipolar patches separated along the curved path of the narrow zone, and sheared with each other and also with the previously emerged bipoles (from panel (e) to panel (f)).

\begin{table}
\begin{center}
\caption{List of all the 4 X-class and 27 M-class flares in AR 12673.} \label{tab:1}
\centering
\begin{tabular}{ccccccc}
\\
\tableline\tableline
Date & GOES & \multicolumn{3}{c}{Time (UT)} & Duration  \\
\cline{3-5}
(UT) & Class & Start & Peak & End & (minutes)   \\
\tableline
2017 Sep ~4 & M1.2 & 05:36 & 05:49 & 06:05 & 29  \\
2017 Sep ~4 & M1.5 & 15:11 & 15:30 & 15:33 & 22  \\
2017 Sep ~4 & M1.0 & 18:05 & 18:22 & 18:31 & 26  \\
2017 Sep ~4 & M1.7 & 18:46 & 19:37 & 19:52 & 66  \\
2017 Sep ~4 & M1.5 & 19:59 & 20:02 & 20:06 & 7  \\
2017 Sep ~4 & M5.5 & 20:28 & 20:33 & 20:37 & 9  \\
2017 Sep ~4 & M2.1 & 22:10 & 22:14 & 22:19 & 9  \\
2017 Sep ~5 & M4.2 & 01:03 & 01:08 & 01:11 & 8  \\
2017 Sep ~5 & M1.0 & 03:42 & 03:51 & 04:04 & 22  \\
2017 Sep ~5 & M3.2 & 04:33 & 04:53 & 05:07 & 34  \\
2017 Sep ~5 & M3.8 & 06:33 & 06:40 & 06:43 & 10  \\
2017 Sep ~5 & M2.3 & 17:37 & 17:43 & 17:51 & 14  \\
2017 Sep ~6 & X2.2 & 08:57 & 09:10 & 09:17 & 20  \\
2017 Sep ~6 & X9.3 & 11:53 & 12:02 & 12:10 & 17  \\
2017 Sep ~6 & M2.5 & 15:51 & 15:56 & 16:03 & 12  \\
2017 Sep ~6 & M1.4 & 19:21 & 19:30 & 19:35 & 14  \\
2017 Sep ~6 & M1.2 & 23:33 & 23:39 & 23:44 & 11  \\
2017 Sep ~7 & M2.4 & 04:59 & 05:02 & 05:08 & 9  \\
2017 Sep ~7 & M1.4 & 09:49 & 09:54 & 09:58 & 9  \\
2017 Sep ~7 & M7.3 & 10:11 & 10:15 & 10:18 & 7  \\
2017 Sep ~7 & X1.3 & 14:20 & 14:36 & 14:55 & 33  \\
2017 Sep ~7 & M3.9 & 23:50 & 23:59 & 00:14 & 24  \\
2017 Sep ~8 & M1.3 & 02:19 & 02:24 & 02:29 & 10  \\
2017 Sep ~8 & M1.2 & 03:39 & 03:43 & 03:45 & 6  \\
2017 Sep ~8 & M8.1 & 07:40 & 07:49 & 07:58 & 18  \\
2017 Sep ~8 & M2.9 & 15:09 & 15:47 & 16:04 & 55  \\
2017 Sep ~8 & M2.1 & 23:33 & 23:45 & 23:56 & 23  \\
2017 Sep ~9 & M1.1 & 04:14 & 04:28 & 04:43 & 29  \\
2017 Sep ~9 & M3.7 & 10:50 & 11:04 & 11:42 & 52  \\
2017 Sep ~9 & M1.1 & 22:04 & 23:53 & 00:41 & 157  \\
2017 Sep 10 & X8.2 & 15:35 & 16:06 & 16:31 & 56  \\
\tableline
\end{tabular}
\end{center}
\end{table}

\begin{figure*}
\centering
\includegraphics
[bb=45 255 545 595,width=0.8\textwidth]{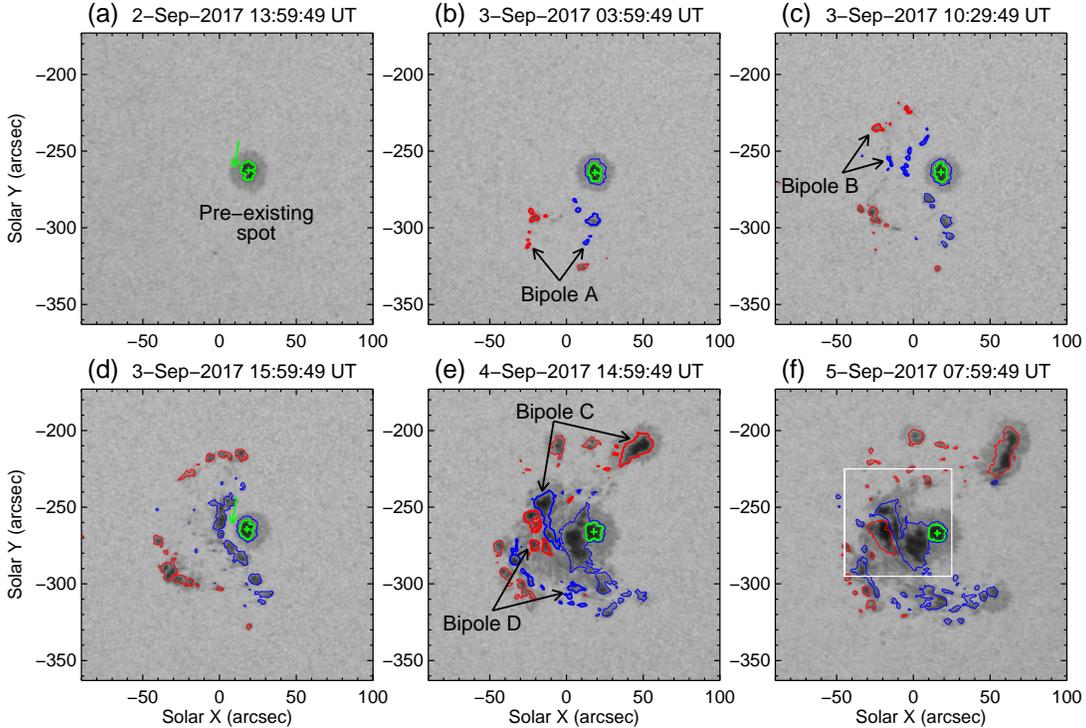} \caption{Sequence of HMI intensity maps displaying the evolution of AR 12673. The green curves and ``+" symbols mark the umbral boundary and intensity centroid of the pre-existing spot, respectively. The blue and red contours represent the LOS magnetic fields at +800 G and -800 G, respectively, among which the bold ones outline the newly emerging bipolar patches. The arrows in panels (a) and (d) denote the penumbra which disappeared due to the conflict with the nearby moving patches. The box in panel (f) outlines the FOV of Figures 3(b)-(e3). \protect\\An animation (Movie1.mp4) of this figure is available. \label{fig1}}
\end{figure*}

\begin{figure*}
\centering
\includegraphics
[bb=43 167 538 669,width=0.9\textwidth]{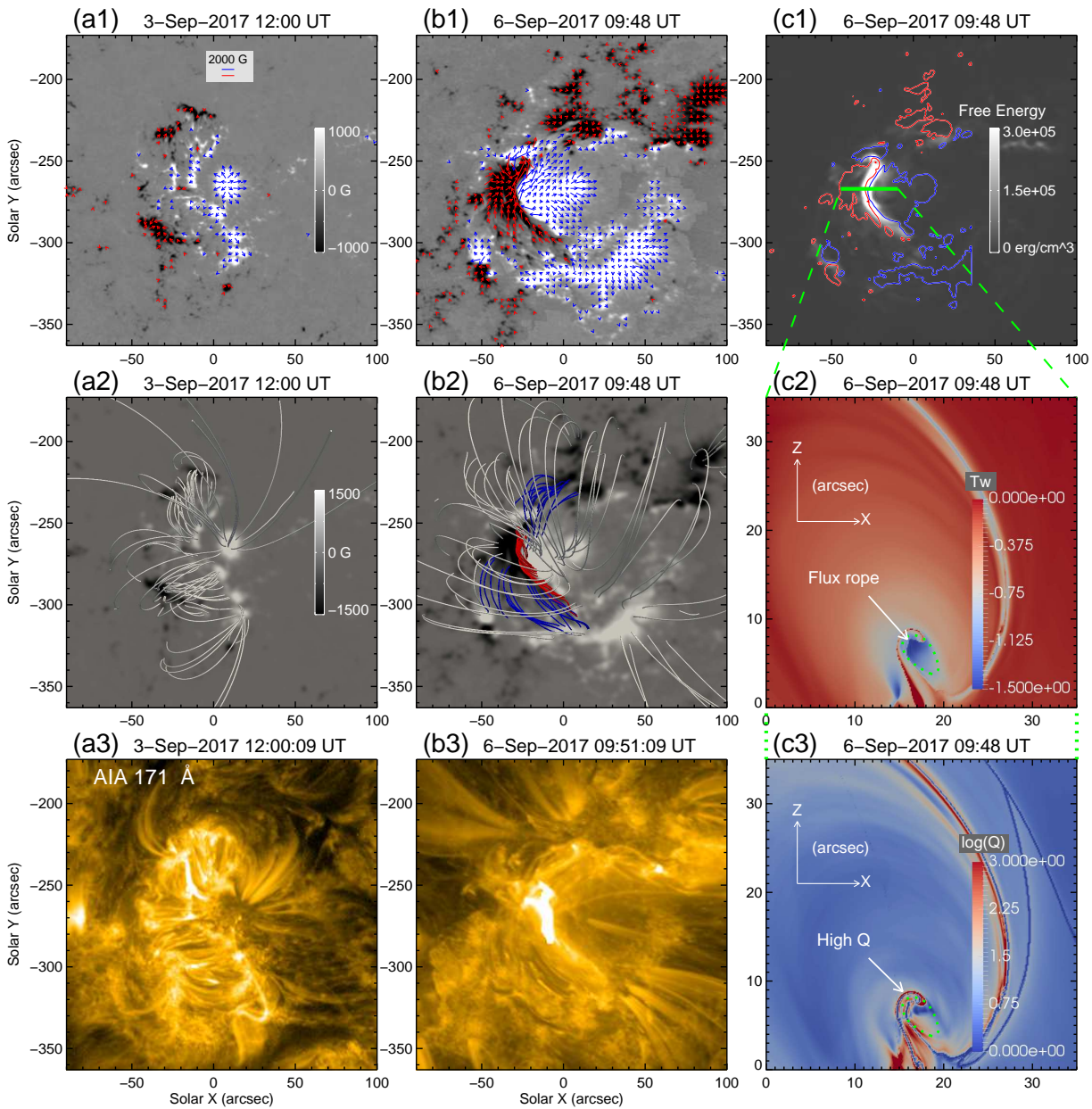} \caption{Panels (a1)-(a3): HMI vector magnetograms at 12:00 UT on Sept. 3, top-view of the extrapolated field lines, and corresponding AIA 171 {\AA} image, respectively. Panels (b1)-(b3): similar to panels (a1)-(a3), but for the time at 09:48 UT on Sept. 6. Panel (c1): free energy density corresponding to panel (b1) overlaid with the vertical magnetic field contours at $\pm$800 G. Panels (c2) and (c3): $\mathcal{T}_w$ and $Q$ distribution in the $x$-$z$ plane along the cut labeled in panel (c1). In panel (b2), the blue field lines connect the opposite patches of bipole ``C" and bipole ``D" (labeled in Figure 1), respectively, and the red field lines indicate a flux rope along the PIL. In panels (c2) and (c3), the green dotted curves outline the general shape of the flux rope. \label{fig2}}
\end{figure*}

\begin{figure*}
\centering
\includegraphics
[bb=46 169 540 674,width=0.85\textwidth]{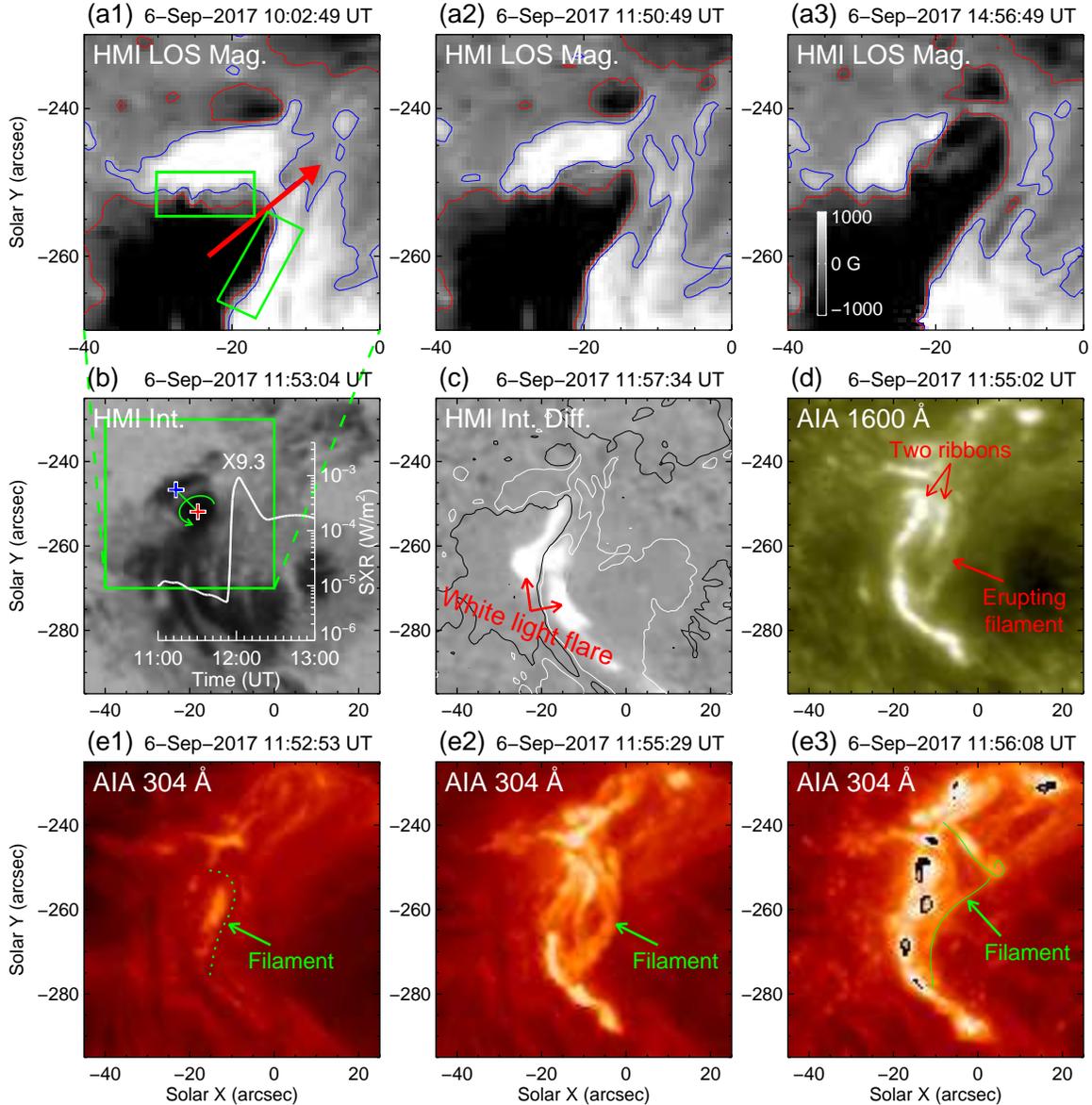} \caption{Panels (a1)-(a3): HMI LOS magnetograms with contours of $\pm$300 G showing the rapid shearing motion around the X9.3 flare. The red arrow in panel (a1) denotes the moving direction of the negative patch, and the rectangles outline two areas with rapid shearing motion. Panels (b) and (c): HMI intensity map at 11:53:04 UT on Sept. 6 and the intensity difference map from that of 3.5 min later overlaid with the LOS magnetic field contours. The red and blue ``+" symbols in panel (b) mark the intensity centroids of two opposite polarity patches, and the green curve with arrow denotes the rotation of the negative patch. The inserted white curve displays the \emph{GOES} soft-X-ray flux variation of the X9.3 flare. Panel (d): AIA 1600 {\AA} image showing the erupting filament and the associated two ribbons. Panels (e1)-(e3): Sequence of AIA 304 {\AA} images showing the eruption process of the filament. \protect\\Two animations (Movie2.mp4 \& Movie3.mp4) of this figure are available. \label{fig3}}
\end{figure*}

\begin{figure*}
\centering
\includegraphics
[bb=120 302 490 600,clip,angle=0,width=0.85\textwidth]{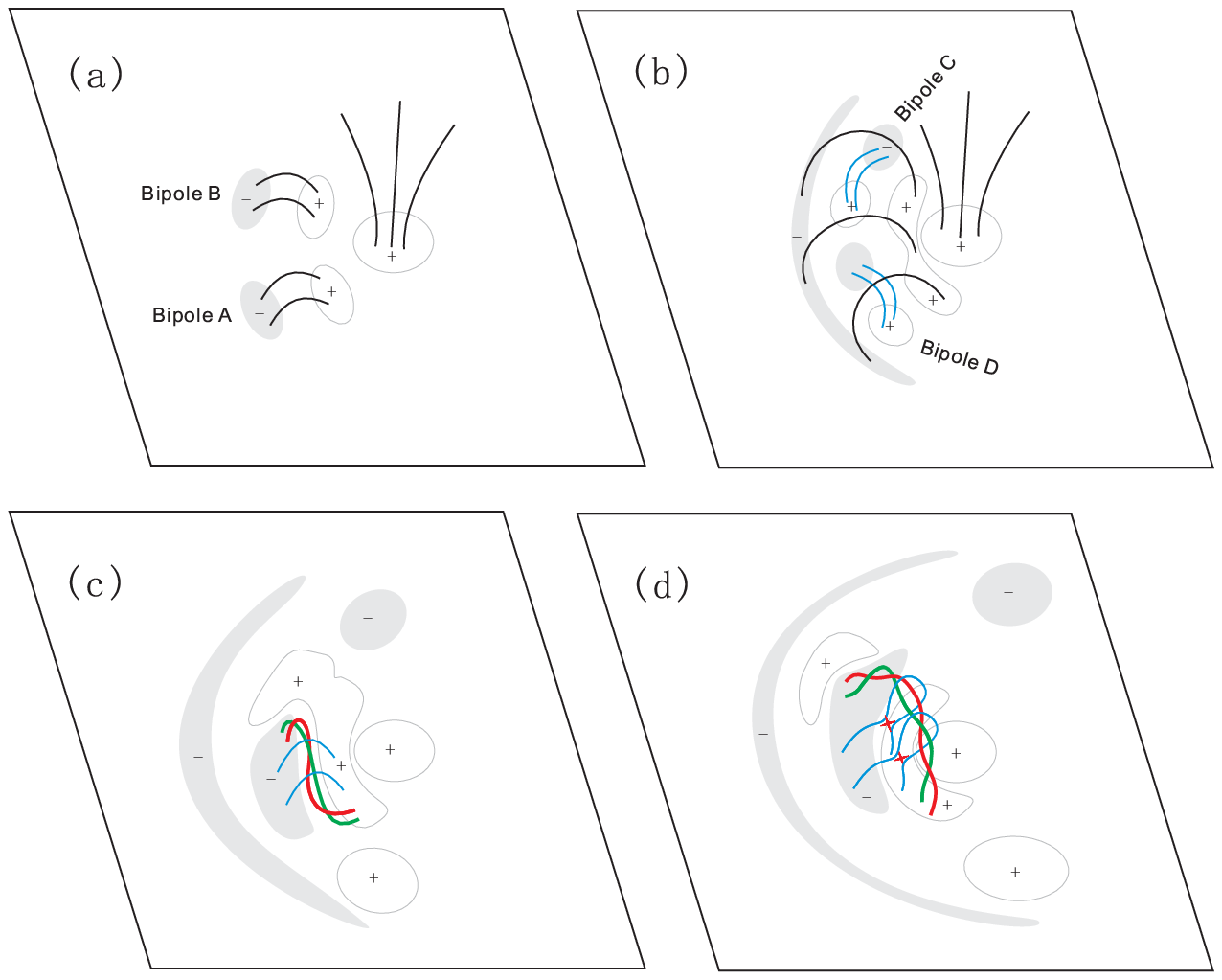} \caption{Cartoon model illustrating the block-induced eruption. In panel (a), the right-most patch represents the pre-existing spot with the positive polarity, and the two pairs of patches at its left side connecting with field lines are two emerging bipoles ``A" and ``B". In panel (b), the patches connecting with blue lines are two newly emerging bipoles ``C" and ``D" in the narrow zone between the oppositive polarities of bipoles ``A+B". In panel (c), the braiding red and green lines form a flux rope and the blue lines represent the overlying loops. In panel (d), the red cross symbols mark the reconnection sites between the anti-directed field lines of the stretched overlying loops due to the filament eruption. \label{fig4}}
\end{figure*}

At the early stage of the AR evolution, the magnetic structures of the AR exhibited in the observed vector magnetogram were relatively simple (Figure 2(a1)). The extrapolated coronal field lines show that most loops connect the positive fields on the west with the negative fields on the east (panel (a2)). The extrapolated lines are well consistent with the observed loops in the AIA 171 {\AA} image (panel (a3)). However, the AR magnetic topologies on Sept. 6 were quite complex. The vector magnetogram in panel (b1) reveals that the magnetic gradient between the negative patch of bipole ``D" (labeled in Figure 1) and the encountering positive fields is very steep, and the magnetic fields nearby the curved PIL are highly sheared. From the extrapolated data (panel (b2)), we can see that, there is a flux rope structure outlined by the red field lines above the PIL. The blue field lines in panel (b2) connect the opposite polarity patches of bipoles ``C" and ``D" (labeled in Figure 1), respectively, which coincide with the loop structures in AIA 171 {\AA} image (panel (b3)). Moreover, the distribution of the magnetic free energy calculated with the vector field observation at 09:48 UT on Sept. 6 is displayed in panel (c1). It is clear that the high free energy area (bright region in panel (c1)) is located along the PIL of the opposite polarities. Along the cut marked by the green line, we examine the $\mathcal{T}_w$ and $Q$ distribution above the sheared field. In the $\mathcal{T}_w$ map, there is an ellipse region (outlined by the green dotted curve in panel (c2)) with the high twist number. The high $\mathcal{T}_w$ region is consistent with the flux rope outlined by the red lines in panel (b2). The edge of the high twist region coincides with the high $Q$ as shown in panel (c3). For a magnetic flux rope, the field lines wrapping around the central axis have a similar connectivity, indicating that there indeed exists a flux rope along the PIL.

The photospheric flow went on shearing the magnetic fields near the PIL, as displayed in Figures 3(a1)-(a3) and the associated Movie2. The large negative patch moved north-west (indicated by the red arrow in panel (a1)), and continuously sheared with the surrounding positive fields. Two areas with rapid shearing motion are outlined by two green rectangles. The negative patch intruded into the positive one and split it into two parts (from panels (a2)-(a3)). In the HMI intensity map (panel (b)), the orientation from the intensity centroid (marked by the red ``+") of the negative patch to that (marked by the blue ``+") of the positive patch increased anti-clockwise, due to the shearing motion. In addition, the negative patch itself also rotated anti-clockwise, as denoted by the green curve with arrow (see Movie2). During this process, the X9.3 flare took place, as revealed by the variation of \emph{GOES} soft X-ray flux (see the inserted white curve in panel (b)). When the flare occurred, a white light flare with two ribbons can be identified in the HMI intensity map, which is more prominent in the intensity difference image (panel (c)). The two ribbons were also observed in AIA 1600 {\AA} image (panel (d)). Moreover, an erupting filament was captured by AIA in multi-wavelengths, such as 1600 {\AA}, 304 {\AA}, 131 {\AA}, and 94 {\AA} (see Movie3). Panels (e1)-(e3) show the filament eruption in 304 {\AA}. The filament (outlined by the dotted curve in panel (e1)) was mainly located along the PIL at 11:52:53 UT just before the occurrence of the X9.3 flare, and then it was brightened during its rising motion (panel (e2)). At 11:56:08 UT, the filament appeared as a kink shaped structure, as indicated by the knotted curve in panel (e3).

\section{CONCLUSIONS AND DISCUSSION}

Using the HMI and AIA data, we have studied AR 12673 which erupted 31 X- and M-class flares and numerous lower class ones, including the largest one in Solar Cycle 24. We find that two bipolar regions emerged successively nearby the only one pre-existing spot. The movement of the emerging patches was blocked by the pre-existing spot and consequently distorted and elongated, forming two semi-circular shaped structures. Then within the semi-circular narrow region, two sequences of new bipoles emerged and separated along the curved channel. The new bipoles sheared and interacted with the surrounding fields, during which numerous flares occurred. At the PIL between the newly emerging negative fields and the surrounding positive ones, the magnetic fields were highly sheared, resulting in the accumulation of magnetic free energy. The extrapolated coronal fields and the calculation of the $\mathcal{T}_w$ and $Q$ reveal that there existed a flux rope along the PIL. One negative patch began to rapidly rotate and move forward, increasing the shear at the PIL, and then the X9.3 flare took place. At the early stage of the flare, an erupting filament was observed to rise in multi-wavelength UV and EUV images, and the filament presented a kink motion during its eruption.

At the initial emerging stage of ARs, the magnetic topologies are relatively simple, close to potential fields. However, the photospheric flows can greatly shear the magnetic fields and make them highly non-potential. The free energy will be stored in the non-potential field and can be used to power solar flares, and CMEs (Schrijver et al. 2008; Wang et al. 2015). In the present study, AR 12673 evolved from a simple spot to a complex system accompanied by numerous flares. We sketch a cartoon model to illustrate the formation of the complex structures, as shown in Figure 4. At first, there only existed one sunspot (represented by the right-most ellipse in panel (a)), and subsequently two pairs of bipolar regions ``A" and ``B" successively emerged at its left. Then the movement of the bipolar fields was blocked by the pre-existing spot, and the distorted shape of the bipoles looks like a crescent shape (panel (b)). Within the narrow zone, bipolar regions ``C" and ``D" began to emerge and separate along the semi-circular curve, and the elongated field lines together with the surrounding lines formed a complex system, responsible for the consequent occurrence of the flares. However, besides the bipole-blocking process, the magnetic flux emergence and cancellation may also play an important role in creating the flare-producing conditions in this AR, similar to the previous studies (e.g., Zirin \& Wang 1993; Zhang et al. 2001; Louis et al. 2015). We can see that a great deal of magnetic flux (bipoles ``A", ``B", ``C", and ``D" shown in Figure 1) emerged continuously to the east of the pre-existing sunspot. Consequently, flux cancellation among these bipoles took place at some interaction areas (see Movie1), which may contribute to the flare-productivity.

For two-ribbon flares and CMEs, Amari et al. (2000) proposed a model in which twisted flux ropes representing the magnetic structures of filaments play a crucial role. The erupting flux ropes stretch the overlying loops, and the reconnection takes place between the anti-directed lines below, resulting in solar flares (Shibata et al. 1995; Lin \& Forbes 2000). The flux rope in this study was located above the strong PIL shown in Figure 2, and its two ends were anchored in the negative (on the left of the PIL; the negative patch of bipole ``D") and positive (right; merged from the positive fields of bipoles ``C", ``A", and ``B", instead of ``D") polarities. The opposite polarity patches on the two sides of the PIL were not initially connected, and their connection was formed during the following interactions. Therefore, the flux rope seems to not emerge from below. For the formation of flux ropes, the shearing motion and rotation of sunspots are found to play an important role (e.g., Yan et al. 2015). In addition, magnetic diffusion/flux-cancellation processes at the PIL were crucial for the formation of this flux rope, e.g., illustrated by the tether-cutting model (Moore et al. 2001). As proposed by Moore et al. (2001), the sheared magnetic arcades reconnect at their inner legs, and then a longer twisted flux rope can be formed over the PIL.

There are some different mechanisms for the onset of filament eruptions, such as  magnetic breakout (Antiochos et al. 1999), tether-cutting (Moore et al. 2001), kink instability (T{\"o}r{\"o}k \& Kliem 2005), and torus instability (Kliem \& T{\"o}r{\"o}k 2006). Based on the observations, we find that the X9.3 flare on Sept. 6 was triggered by an erupting filament due to the kink instability (Figures 3(e1)-(e3)). For a twisted flux rope, the threshold value of kink instability is mainly about $|\mathcal{T}_w|$=1.75 (Torok \& Kleim 2003). The force-free model for the time 6 Sept. 09:48 UT shows the existence of the flux rope above the strong PIL (see Figure 2). The average $\mathcal{T}_w$ of the inner part of the flux rope is about $-$1.5, and that of the outer part is about $-$1.2. After that, the rapid shearing motion and rotation of the related magnetic structures were observed, as shown in Figure 3 (also the associated animation). Consequently, the twist number of the flux rope continuously increased, and ultimately reached or even exceeded the threshold of the kink instability. This process can be illustrated in Figures 4(c)-(d). Due to the shearing motion of opposite polarities, a great deal of free energy was accumulated at the PIL region. Meanwhile, twisted flux rope can be created along the highly sheared PIL. In Figure 4(c), the braiding red and green lines represent a flux rope, corresponding to that in Figure 2. Because of the continuous shearing motion displayed in Figures 3(a1)-(a3), the twist of the flux rope consequently increased. Eventually, the filament was kink unstable and erupted. In Figure 3(e3), the kink knot of the filament during the eruption is the most important observational character of the kink instability. The rising filament stretched the overlying loops (blue lines in Figure 4(d)), and then the stretched field lines reconnected at the sites marked by the red symbols, thus triggering the solar flare.

In the previous studies, there are mainly three types of mechanisms for triggering major flares: (1) flux emergence and cancellation, (2) shearing motions, and (3) sunspot rotation. The block-induced eruption model proposed in this Letter can be considered to be the fourth mechanism. Different from the former three ones, in our model, there exists a standing spot which blocks the horizontal movement of newly emerging patches, resulting in the formation of complex sheared structures. Using this model, we can answer the two questions raised in the first part of this paper: the block-induced complex structures built the flare-productive AR 12673 and the X9.3 flare was triggered by an erupting filament due to the kink instability.

\acknowledgments {We thank the referee for the valuable suggestions and constructive comments. The data are used courtesy of HMI and AIA science teams. This work is supported by the National Natural Science Foundations of China (11673035, 11533008, 11403044, 41404136, 11373004), Key Programs of the Chinese Academy of Sciences (QYZDJ-SSW-SLH050), and the Youth Innovation Promotion Association
of CAS (2014043).\\ }

{}

\clearpage

\end{document}